\documentclass[final,3p,times]{elsarticle}

\usepackage{amssymb}

\usepackage[nodots]{numcompress}
\usepackage{amsmath}
\usepackage{algorithm}
\usepackage{algorithmic}
\usepackage{url}
\usepackage{multicol}
\newcommand{\Tt}[1]{\mathbf{#1}}
\providecommand{\vnorm}[1]{ \lVert #1 \rVert }

\journal{Computers \& Fluids}

\begin{document}

\begin{frontmatter}



\title{Explicit Parallel-in-time Integration of a Linear Acoustic-Advection System}


\author[label1]{D. Ruprecht\corref{cor1}}
\cortext[cor1]{Corresponding author.}
\ead{daniel.ruprecht@usi.ch}

\author[label1]{R. Krause}
\address[label1]{Institute of Computational Science, Universit{\`a} della Svizzera italiana, Via Giuseppe Buffi 13, 6906 Lugano, Switzerland}

\begin{abstract}
The applicability of the Parareal parallel-in-time integration scheme for the solution of a linear, two-dimensional hyperbolic acoustic-advection system, which is often used as a
test case for integration schemes for numerical weather prediction (NWP), is addressed. Parallel-in-time schemes are a possible way to increase, on the algorithmic level, the amount of parallelism, a requirement arising from the rapidly growing number of CPUs in high performance computer systems. A recently introduced modification of the "parallel
implicit time-integration algorithm" could successfully solve hyperbolic problems arising in structural dynamics. It has later been cast into the framework of Parareal. The present
paper adapts this modified Parareal and employs it for the solution of a hyperbolic flow problem, where the initial value problem solved in parallel arises from the spatial discretization of a partial differential equation by a finite difference method. It is demonstrated that the modified Parareal is stable and can produce reasonably accurate solutions while allowing for a noticeable reduction of the time-to-solution. The implementation relies on integration schemes  already widely used in NWP (RK-3, partially split forward Euler, forward-backward). It is demonstrated that using an explicit partially split scheme for the coarse integrator allows to avoid the use of an implicit scheme while still achieving speedup.
\end{abstract}


\end{frontmatter}


\section{Introduction}
Numerical weather prediction (NWP) is a classical application area for high performance computing and the list of the 500 most powerful supercomputers in the world\footnote{\url{www.top500.org}} contains a number of systems operated by weather services and meteorological research centers. As the
growth of the frequency of individual processors has stopped several years ago because of fundamental physical problems, increases in performance
of supercomputers are now achieved by a rapidly growing number of nodes and cores. Contemporary massively parallel computers already feature up to several hundred thousand CPUs. This change towards massively parallel systems requires software to feature increasing levels of parallelism in order to run efficiently on computers
with more and more CPUs, see for example \cite{Sutter2005, AsanovicEtAl2009} for discussions of this point in general and \cite{WehnerEtAl2008} for
a recent assessment specifically for weather and climate simulations and also \cite{DrakeFoster1995} for a much earlier one.

Parallelization by domain decomposition is by now a standard technique employed in NWP codes like for example COSMO\footnote{\url{www.cosmo-model.org}} or WRF\footnote{\url{www.wrf-model.org}}. It relies on the decomposition of the computational domain into subdomains and the assignment of the computations for a particular subdomain to one processor. While this approach can yield very good parallel scaling to a large number of cores, it nevertheless saturates at some critical number of processors beyond which the subdomains become too small. In view of the fact that NWP is a time critical application, additional directions of parallelization are required in order to achieve a maximized reduction of the time-to-solution on upcoming computer systems with quickly growing numbers of cores. One possibility is to modify codes on the implementation level, as done
for example in \cite{Michalakes2008}, where parts of WRF are ported to graphic processing units (GPU) thereby achieving significant acceleration. Investigating numerical
algorithms inherently suitable for parallel computation is another important direction. Parallel-in-time integration schemes, allowing for a parallelization of the time-stepping procedure, are a possible approach to increase parallelism on the algorithmic level.

The Parareal algorithm for parallel-in-time integration has been introduced in \cite{LionsEtAl2001}. It relies on the introduction of a computationally cheap, coarse integrator,
which is used to produce guesses of the solution at several later points in time. Starting from these guesses, an accurate but computationally expensive fine integrator
is run concurrently. The results are then used to propagate a correction to the guesses by again running the coarse integrator serially over all intervals. This procedure is iterated and converges to the solution that would be obtained by running the fine propagator sequentially. In \cite{BalMaday2002} the algorithm is modified and used to solve a nonlinear parabolic PDE arising in financial mathematics. Since then, Parareal has been applied successfully to a broad range of problems. 
In \cite{MadayTurinici2003a} it is applied to problems arising in quantum chemistry, in \cite{Trindade2004, Trindade2006, FischerEtAl2003} to the Navier-Stokes equations,
in \cite{Liu2008} it is used for the Princeton ocean model while in \cite{SamaddarEtAl2010} Parareal is employed to successfully speed up simulations of fully turbulent plasma. Finally,  a hybrid approach coupling Parareal and spectral deferred correction methods is presented in \cite{Minion2010}.

Besides several applications, a number of theoretical results have been published as well. In \cite{Gander2007}, a comprehensive mathematical analysis is provided, including the interpretation of Parareal as a multiple-shooting as well as multigrid-in-time method. Also super-linear convergence on short and linear convergence on long time intervals
is shown. A super-linear convergence theorem for Parareal when applied to nonlinear systems of ODEs is proven in  \cite{GanderHairer2007}.  Convergence for different numerical examples is demonstrated, among others Lorenz equations and a system arising from the discretization of the viscous Burgers equation. Also, NWP is mentioned as a possible application where Parareal could be beneficial.

Inspired by the original Parareal, a parallel-in-time scheme called "parallel implicit time-integration algorithm" (PITA) is introduced in  \cite{FarhatEtAl2003} and its applicability to three model problems from fluid, structure and fluid-structure interaction applications is investigated. While PITA is found to work well for parabolic and first order hyperbolic problems, stability issues for second order hyperbolic systems are identified. Similar instabilities of Parareal applied to hyperbolic equations are indicated in \cite{Bal2003}. Stability criteria for Parareal are derived in \cite{StaffRonquist2003} and it is found that Parareal is unstable for problems with pure imaginary eigenvalues or eigenvalues with a dominant imaginary part and hence that the original version of Parareal is not applicable to flow problems that are strongly dominated by advection. Further analysis of the problems arising from imaginary eigenvalues can be found in \cite{Gander2007}.  Finally, it is shown in \cite{Gander2008} that Parareal can be efficient for advection on bounded domains, where the solution is mainly determined by the boundary values and not by the initial value. It is also shown that in general Parareal is inapplicable to even the one dimensional version
 of the wave equation. These limitations most likely forbid using the original Parareal to solve the equations arising in NWP. Indeed, it is confirmed in the present paper that
 the original Parareal can develop instabilities for the two-dimensional advection as well as the two-dimensional acoustic-advection problem, although the sequentially run coarse and fine integrator both remain stable.
 
However, a modification of PITA is introduced in \cite{FarhatCortial2006} that is demonstrated to work for second-order hyperbolic problems arising in linear structural dynamics. It
is extended and applied to nonlinear problems in \cite{FarhatCortial2008}. In \cite{GanderPetcu2009} it is shown that for linear problems PITA and Parareal are
equivalent, the new version of PITA is cast into the Parareal framework and demonstrated to successfully solve second order ODEs. Also the name "Krylov-subspace-enhanced Parareal" is coined, which is adopted in the present paper and will be abbreviated KSE-Parareal from now on. To the authors' knowledge, no works exist that address the feasibility of KSE-Parareal for hyperbolic flow problems. Also, the results obtained for the examples from structural dynamics relied on the use of an implicit coarse propagator in order to allow for large coarse time steps. This poses a problem when considering a possible application of KSE-Parareal for NWP: While there are codes where implicit solvers are used to model atmospheric flows, see for example \cite{PrusaEtAl2008}, most NWP codes rely on explicit integration schemes. Adding an efficient nonlinear solver into such an existing code would be at best a challenging endeavor.

As a first step to investigate the feasibility of using KSE-Parareal for NWP, the present paper investigates the applicability to a linear two-di\-men\-sio\-nal acoustic-advection system. This system is often used as a simplified test case for feasibility or stability studies of integration schemes for NWP applications, see for example \cite{SkamarockKlemp1992, WickerSkamarock2002, Gassmann2005, Baldauf2010}.  It is demonstrated that for this type of problem an explicit partially split scheme (see 
\cite{SkamarockKlemp1992, WickerSkamarock2002, WickerSkamarock1998}) with strong divergence damping can be employed as coarse propagator, avoiding the necessity for the implementation of an implicit scheme. Partially split methods are widely used in codes for numerical weather prediction and a possible implementation of parallel-in-time integration could rely on these already implemented schemes (compare for the comments in subsection \ref{subsec:originalParareal}). Both efficiency and accuracy of the parallel-in-time scheme for the investigated example problem are discussed. A comparison with the speedup obtained by switching to a partially split scheme is given and it is found that KSE-Parareal yields larger speedup while maintaining comparable accuracy.

The present paper is motivated by addressing the potential feasibility of applying KSE-Parareal in codes for NWP and the choice of parameters and methods is inspired
by this goal. However, the investigated test problem, as a reduced model emerging from the Euler equations, is of potential interest to a much broader field of applications.
It contains the two-dimensional wave equation as a special case, which is a model arising in numerous applications. It is also mathematically equivalent to the linearized shallow water equations, which are another popular reduced model in atmospheric as well as in ocean sciences.

\section{The Parareal Algorithm}\label{sec:algorithm}
This section recapitulates the Parareal parallel-in-time algorithm, first in subsection \ref{subsec:originalParareal} in
its original version and then in subsection \ref{subsec:KSEparareal} in the modified version, suitable for hyperbolic
systems.

\subsection{Original Version}\label{subsec:originalParareal}
The presentation of Parareal in this section follows the formulation of the algorithm as a predictor-corrector method 
in \cite{Gander2007}. The aim is to solve an initial value problem of the type
\begin{equation}
	\label{eq:ODE}
	\Tt{q}_{t} = \Tt{f}(\Tt{q}), \quad \Tt{q}(0) = \Tt{q}_{0} \in \mathbb{R}^{d}, \quad t \in [0, T]
\end{equation}
in parallel. In the following, $\Tt{f}$ stems from the spatial discretization of a partial differential equation ("method-of-lines") and $d$ is the total number of
degrees-of-freedom. Let $\mathcal{F}_{\delta t}$ denote some integration scheme of appropriate accuracy, using a constant time step
$\delta t$. In order to employ Parareal, a second integration scheme $\mathcal{G}_{\Delta t}$ is required, which has to be 
much cheaper than $\mathcal{F}_{\delta t}$ in terms of computation time but which can also be much less accurate. Usually,
$\mathcal{G}_{\Delta t}$ will employ a larger time step, that is $\Delta t \gg \delta t$, and be of lower order, so that less evaluations of the right hand
side $\Tt{f}$ are required. Further, $\mathcal{G}_{\Delta t}$ can also feature a less accurate spatial discretization, using
lower order spatial operators and/or a coarser spatial mesh (see for example \cite{FischerEtAl2003}), leading to a different right hand side function
$\Tt{f}_{\rm c} \neq \Tt{f}$ for the case of a semi-discrete PDE. As demonstrated in \cite{KaberMaday2007}, the coarse propagator
can even be based on simplified or averaged model physics. Following common terminology, $\mathcal{F}_{\delta t}$ will be referred
to as the fine propagator and $\mathcal{G}_{\Delta t}$ as the coarse propagator.

For a given $\Delta t$ assume that the time interval $[0, T]$ can be decomposed into $M_{\rm c}$ coarse intervals with 
endpoints
\begin{equation}
  0 = t_{0} < t_{1} < \ldots < t_{M_{\rm c}} = T
\end{equation}
and constant lengths
\begin{equation}
  \Delta t = t_{i+1} - t_{i}, \quad i=0,\ldots,M_{\rm c}-1.
\end{equation}
Further let $\Delta t$ be a multiple of the fine time step $\delta t$, that is there exists an integer 
$N_{\rm f} \in \mathbb{N}$ such that
\begin{equation}
  \Delta t = N_{\rm f} \delta t
\end{equation}
and every coarse interval contains $N_{\rm f}$ fine intervals, hence
\begin{equation}
  T = M_{\rm c} \Delta t = \left( M_{\rm c} N_{\rm f} \right) \delta t.
\end{equation}
Finally assume that $M_{\rm c}$ is a multiple of the number of available processors $N_{\rm p}$ so that integration from
$t=0$ to $t=T$ can be done in $M_{\rm p} := M_{\rm c}/N_{\rm p}$ parallel steps of length $N_{\rm p} \Delta t$, each step distributing
$N_{\rm c} := M_{\rm c}/M_{\rm p} = N_{\rm p}$ coarse intervals onto $N_{\rm p}$ processors. Note that in principal one can
also assign multiple coarse intervals to a single processor, that is let $N_{\rm c}$ be a multiple of $N_{\rm p}$, in order to perform larger parallel steps but
below it is always assumed that $N_{\rm c} = N_{\rm p}$. The number of fine steps per parallel step is equal to
$N_{\rm t} := N_{\rm c} N_{\rm f}$. 

Denote by
\begin{equation}
  \mathcal{F}_{\delta t}(\Tt{q}, t_{1}, t_{0}), \quad \mathcal{G}_{\Delta t}(\Tt{q}, t_{1}, t_{0})
\end{equation}
the result of the fine or coarse propagator if used to integrate \eqref{eq:ODE} from an initial value $\Tt{q}$ at time $t_{0}$ to
a time $t_{1} > t_{0}$ using time steps $\delta t$ or $\Delta t$ respectively. Let a subscript $n$ denote the approximation of $\Tt{q}$ at time $t^{n}$, that is
\begin{equation}
	\Tt{q}_{n} \approx \Tt{q}(t^{n})
\end{equation}
and let a superscript $k$ denote quantities in the $k$-th iteration of Parareal. With this notation, the basic iteration of Parareal performed in one 
parallel step reads
{\newline
\parbox{0.35\textwidth}{\vspace{1ex}
\begin{align*}
	\Tt{q}^{k+1}_{n+1} =& \mathcal{G}_{\Delta t} \left(\Tt{q}^{k+1}_{n} , t_{n+1}, t_{n} \right) 
        + \mathcal{F}_{\delta t} \left( \Tt{q}^{k}_{n}, t_{n+1}, t_{n} \right) \\
		& - \mathcal{G}_{\Delta t} \left( \Tt{q}^{k}_{n} , t_{n+1}, t_{n} \right),
\end{align*} }\hfill
\parbox{0.001\textwidth}{\begin{align} \label{eq:parareal}  \end{align}} \newline}
compare for \cite{Gander2007}. For $k \to N_{\rm c}$, the iteration converges to a a solution $\Tt{q}_{n}$, $n=0,\ldots, N_{\rm c}$, satisfying
\begin{equation}
	\Tt{q}_{n+1} = \mathcal{F}_{\delta t} \left(  \Tt{q}_{n}, t_{n+1} , t_{n}  \right),
\end{equation}
that is to an approximation of the exact solution with the accuracy of the fine propagator. After $N_{\rm c}$ iterations, the algorithm will 
always provide the sequential solution, but in order to be efficient, it has to converge after $N_{\rm it} \ll N_{\rm c}$ iterations. Note that as the 
values $\Tt{q}^{k}_{n}$ in \eqref{eq:parareal} are given from either the initialization or the previous iteration, the expensive computations of the values 
$\mathcal{F}_{\delta t}(\Tt{q}^{k}_{i}, t_{i+1}, t_{i})$ in \eqref{eq:parareal} are all independent and can be performed concurrently.
\begin{algorithm}[t]
\begin{algorithmic}[1]
\STATE \COMMENT{Initialization:}
\STATE $\Tt{q}^{0}_{0} = \Tt{q}_{0}$
\FOR{$i=0$ to $N_{\rm c}-1$}
	\STATE $\Tt{q}^{0}_{i+1} = \mathcal{G}_{\Delta t}(\Tt{q}^{0}_{i}, t_{i+1}, t_{i})$
\ENDFOR
\STATE \COMMENT{Iteration:}
\STATE $k:=0$
\REPEAT
	\STATE \COMMENT{Parallel predictor step:}
	\FOR{$i=0$ to $N_{\rm c}-1$} \label{line:loopStart}
		\STATE $\tilde{\Tt{q}}^{k}_{i+1} = \mathcal{F}_{\delta t}(\Tt{q}^{k}_{i}, t_{i+1}, t_{i})$ \label{line:fineProp}
	\ENDFOR \label{line:loopEnd}
	\STATE  \COMMENT{Sequential correction step:}
	\FOR{$i=0$ to $N_{\rm c}-1$}
	\STATE $\Tt{q}^{k+1}_{i+1} = \mathcal{G}_{\Delta t}(\Tt{q}^{k+1}_{i}, t_{i+1}, t_{i}) + \tilde{\Tt{q}}^{k}_{i+1} - \mathcal{G}_{\Delta t}(\Tt{q}^{k}_{i}, t_{i+1}, t_{i})$ \label{line:Correction}
	\ENDFOR
	\STATE $k := k+1$
\UNTIL{$k=N_{\rm maxit}$}
\end{algorithmic}
\caption{Original version of Parareal}\label{alg:parareal}
\end{algorithm}

Algorithm \ref{alg:parareal} sketches the complete iteration for one parallel step in pseudo code. For simplified notation, the
endpoints of the involved $N_{\rm c}$ coarse intervals have been re-indexed to $t_{0}, \ldots , t_{N_{\rm c}}$. If implementations of
the propagators $\mathcal{F}_{\delta t}$ and $\mathcal{G}_{\Delta t}$ are available, implementing Parareal does not require much more than implementing this iteration,
although achieving efficiency might require modifications of the implementation of $\mathcal{F}_{\delta t}$ and $\mathcal{G}_{\Delta t}$.
Parallelization of the loop computing the values of the fine propagator can easily be done, for example, by adding the corresponding
OpenMP directives\footnote{\url{www.openmp.org}}. This makes Parareal a good candidate for a hybrid spatial/temporal parallelization approach,
where spatial subdomains are assigned to computational nodes containing several processors that share the same memory.
Communication between subdomains would then be implemented by message-passing between nodes instead of cores while
integration inside a subdomain is performed in parallel employing the cores available inside the node.  It is stressed, however, that 
the algorithm is inherently parallel by design and not limited to a specific model of parallelization. A combination of Parareal with domain decomposition is explored in 
\cite{MadayTurinici2003b}. Also note that only the loop in the lines \ref{line:loopStart}--\ref{line:loopEnd} of Algorithm \ref{alg:parareal} has to be parallelized, so over the rest of the computation the idle CPUs could be assigned to other tasks.

Using a stopping criterion in \ref{alg:parareal} instead of a fixed number of iterations is possible as well by defining an error estimate of the form
\begin{equation}
	r^{k} = \max_{i=1,\ldots,N_{\rm c}} \vnorm{ \Tt{q}^{k+1}_{i} - \Tt{q}^{k}_{i} }.
\end{equation}
The iteration would then be terminated if either $k$ reaches the maximum number of iterations allowed or if the error estimate
drops below some prescribed threshold. Throughout this paper, to allow for easier comparison of results, a fixed number of iterations is
carried out without considering an error estimate. 

\subsection{Krylov-subspace-enhanced Parareal}\label{subsec:KSEparareal}
While the original version of Parareal presented in subsection \ref{subsec:originalParareal} has been used successfully for a large number of different types of problems, it was found for example in \cite{Gander2007, FarhatEtAl2003, StaffRonquist2003} that instabilities can arise if it is applied to hyperbolic problems. This subsection presents the modified version of PITA/Parareal introduced in \cite{FarhatCortial2006, FarhatCortial2008}
which proved to be stable when used to integrate hyperbolic problems in structural dynamics and which is applied to a hyperbolic flow problem in the present paper. The presentation given here is different from the original one in \cite{FarhatCortial2006, FarhatCortial2008} and follows again the interpretation of the algorithm in the Parareal framework in \cite{GanderPetcu2009}.

In the original version of Parareal sketched in Algorithm \ref{alg:parareal}, the results $\tilde{\Tt{q}}^{n}_{i+1}$ from the fine propagator computed in line \ref{line:fineProp} are used in the correction step in line \ref{line:Correction} and then thrown away. The idea of the modified version is to keep this information in order to successively enhance the coarse propagator by reusing values from the fine integrator computed in previous iterations. Define by
\begin{equation}
	\Tt{S}^{k} := \textrm{span} \left\{ \Tt{q}^{j}_{i} : i=0,\ldots,N_{\rm c}-1, \quad j=0,\ldots k \right\}
\end{equation}
the subspace spanned by all values on the coarse mesh from previous iterations. Further, denote by $\mathcal{F}(\Tt{S}^{k})$ 
the space spanned by the results obtained by applying the fine propagator to these values, that is by
\begin{equation}
	\Tt{\tilde{q}}^{j}_{i+1} = \mathcal{F}_{\delta t}(\Tt{q}^{j}_{i}, t_{i+1}, t_{i}), \ i=0,\ldots,N_{\rm c}-1, \ j=0,\ldots,k.
\end{equation}
Finally, let $\Tt{P}^{k}$ be the orthogonal projection onto $\Tt{S}^{k}$ with respect to the standard Euclidean scalar product. The coarse propagator is now replaced by 
\begin{equation}
	\label{eq:modifiedCoarse}
	\mathcal{K}_{\Delta t}(\Tt{q}, t_{i+1}, t_{i}) := \mathcal{G}_{\Delta t}( (\Tt{I} - \Tt{P}^{k}) \Tt{q}, t_{i+1}, t_{i}) 
        + \mathcal{F}_{\delta t}(\Tt{P}^{k} \Tt{q}, t_{i+1}, t_{i})
\end{equation}
resulting in Algorithm \ref{alg:KSEParareal}. By definition $\Tt{P}^{k} \Tt{q} \in \Tt{S}^{k}$, so letting $\Tt{s}_{1}, \ldots , \Tt{s}_{r}$ denote an
orthogonal basis of $\Tt{S}^{k}$, for a linear problem $\mathcal{F}_{\delta t}(\Tt{P}^{k}\Tt{q}, t_{i+1}, t_{i})$ can be computed from
{\newline
\parbox{0.35\textwidth}{
\begin{align*}
	\mathcal{F}_{\delta t}( \Tt{P}^{k} \Tt{q}, t_{i+1}, t_{i}) &= \mathcal{F}_{\delta t}\left( \sum_{i=1}^{r} \alpha_{i} \Tt{s}_{i} , t_{i+1}, t_{i} \right) \\
		&= \sum_{i=1}^{r} \alpha_{i} \mathcal{F}_{\delta t}( \Tt{s}_{i}, t_{i+1}, t_{i})
\end{align*} }\hfill
\parbox{0.001\textwidth}{ \begin{align}\label{eq:Genhance} \end{align} } \newline}
where the $\mathcal{F}_{\delta t}(\Tt{s}_{i}, t_{i+1}, t_{i})$ are known from $\mathcal{F}(\Tt{S}^{k})$ and are thus available without running the fine propagator again. 
The information generated in previous iterations successively increases the accuracy of the coarse propagator $\mathcal{K}_{\Delta t}$. As the dimension of $\Tt{S}^{k}$ increases, formally $\Tt{P}^{k} \to \Tt{I}$ and $\mathcal{K}(\Tt{q}) \to \mathcal{F}(\Tt{q})$, see \cite{GanderPetcu2009}. It is also shown there that the method converges once the dimension of $\Tt{S}^{k}$ is no longer increasing. For the nonlinear case the update step becomes more complex, see \cite{FarhatCortial2008} for the nonlinear version of PITA, and no formulation in the Parareal framework is available, yet. Additionally, the nonlinear version relies on using an implicit scheme for $\mathcal{G}_{\Delta t}$ and the adaption for an explicit scheme is not straightforward.

In order to perform the update step $\Tt{S}^{k-1} \rightarrow \Tt{S}^{k}$ in line \ref{line:updateStep} of Algorithm \ref{alg:KSEParareal}, an orthogonal basis of the space
\begin{equation}
	\label{eq:subspaceUpdate}
	\Tt{S}^{k-1} \cup \left\{ \Tt{q}^{k}_{0} , \ldots , \Tt{q}^{k}_{N_{\rm c}-1} \right\}
\end{equation}
has to be computed. In principle this can be done by using the Gram-Schmidt algorithm, keeping the basis of $\Tt{S}^{k-1}$ while successively adding new orthogonal basis vectors. However, the Gram-Schmidt algorithm is known to be unstable, see \cite{Trefethen1997}, and applying it in the context of Parareal does indeed require frequent
re-orthogonalization\footnote{Martin J. Gander and Stefan G\"uttel, personal communication}. Hence in the present paper the update step is performed by computing a full QR decomposition in every step using the LAPACK routine "DGEQP3" \footnote{\url{www.netlib.org/lapack}}. There are highly optimized implementations of this library available for basically every computer architecture and the examples presented in section \ref{sec:results} show that the run time spent in the subspace update is small compared to the time spent for the coarse and fine integrator (see especially figure \ref{fig:percentageRuntime}). Nevertheless, the subspace update does contribute to the sequential part of the algorithm, thus further restricting scalability. Also note that the update of the subspace in line \ref{line:updateStep} introduces a "synchronization point", so that overlapping computation as used in \cite{Minion2010, Aubanel2011} for the original version is not possible here.
\begin{algorithm}[t]
\begin{algorithmic}[1]
	\STATE \COMMENT{Initialization:}
	\STATE $\Tt{y}^{0}_{0} = \Tt{q}_{0}$, $S^{0} = \mathcal{F}(S^{0}) = \emptyset$
	\FOR{$i=0$ to $N_{\rm c-1}$}
	\STATE $\Tt{q}^{0}_{i+1} = \mathcal{G}_{\Delta t}(\Tt{q}^{0}_{i}, t_{i+1}, t_{i})$
	\ENDFOR
	\STATE \COMMENT{Iteration:}
	\STATE $k:=0$
	\REPEAT
	\STATE \COMMENT{Parallel predictor step:}
	\FOR{$i=0$ to $N_{\rm c}-1$}
	\STATE $\tilde{\Tt{q}}^{k}_{i+1} = \mathcal{F}_{\delta t}(\Tt{q}^{k}_{i}, t_{i+1}, t_{i})$
	\ENDFOR
	\STATE \COMMENT{Sequential correction step:}
	\STATE Update $S^{k-1} \rightarrow S^{k}$, $\mathcal{F}(S^{k})$ and $\Tt{P}^{k}$ using  \\ $\Tt{q}^{k}_{i-1}$, $\tilde{\Tt{q}}^{k}_{i}, i=1,\ldots, N_{\rm c}$ \label{line:updateStep}
	\FOR{$i=0$ to $N_{\rm c}-1$}
	\STATE $\Tt{q}^{k+1}_{i+1} = \mathcal{K}_{\Delta t}( \Tt{q}^{k+1}_{i}, t_{i+1}, t_{i})  + \tilde{\Tt{q}}_{i+1}^{k} - \mathcal{K}_{\Delta t}( \Tt{q}^{k}_{i}, t_{i+1}, t_{i})$
	\ENDFOR
	\STATE $k:=k+1$
	\UNTIL{$k=N_{\rm maxit}$}
\end{algorithmic}
\caption{Krylov-subspace-enhanced Parareal}\label{alg:KSEParareal}
\end{algorithm}

\subsection{Expected Speedup}\label{subsec:expectedSpeedup}
The ratio of the sequential to the parallel execution time for a given number of processors is referred to as speedup. Denote by $\tau_{\rm c}$ the time required to
complete one step of length $\Delta t$ of the coarse propagator, by $\tau_{\rm f}$ the time for one step of length $\delta t$ of the fine propagator and by $\tau_{\rm QR}(k)$
the time required to compute the subspace update in iteration $k$.  Recall that $N_{\rm t} = N_{\rm c} N_{\rm f}$ is the total number of $\delta t$-time steps in one
parallel step, $N_{\rm c}$ the total number of $\Delta t$-steps (as defined in subsection \ref{subsec:originalParareal}) and $N_{\rm it}$ the number of iterations. As the dimension of $\Tt{S}^{k}$ increases monotonically with the number of iterations performed \footnote{Note that the QR decomposition is located in the sequential part of the code, so its run
time could probably be reduced by using multithreaded libraries.}, it is
\begin{equation}
	\tau_{\rm QR}(k) \leq \tau_{\rm QR}(N_{\rm it}).
\end{equation}
The speedup obtainable in one parallel step by Algorithm \ref{alg:KSEParareal} using $N_{\rm p}$ processors  can then be estimated by
\begin{align}
	\nonumber
	s(N_{\rm p}) &\approx \frac{ N_{\rm t} \tau_{\rm f}}{ N_{\rm c} \tau_{\rm c} + N_{\rm it} \left(  N_{\rm c} \tau_{\rm c}
			 + \frac{N_{\rm t}}{ N_{\rm p} } \tau_{\rm f}\right) + N_{\rm it}  \tau_{\rm QR}(N_{\rm it})  } \\
		&= \frac{1}{   \left( 1 + N_{\rm it} \right) \left( \frac{N_{\rm c}}{N_{\rm t}} \frac{ \tau_{\rm c} }{ \tau_{\rm f} } \right) 
			+ \frac{N_{\rm it} \tau_{\rm QR}(N_{\rm it})}{N_{\rm t} \tau_{\rm f}}
			+ \frac{ N_{\rm it} }{ N_{\rm p} } }.
	\label{eq:speedup}
\end{align}
See also \cite{Minion2010}. From \eqref{eq:speedup}, three different upper bounds for $s(N_{\rm p})$ can be derived. First, it is
\begin{equation}	
	\label{eq:speedupBound1}
	s(N_{\rm p}) \leq \frac{N_{\rm p}}{N_{\rm it}}
\end{equation}
and hence the efficiency of the algorithm depends critically on convergence within few iterations. Also note that according to \eqref{eq:speedupBound1} the 
parallel efficiency is bounded by $1 / N_{\rm it}$, so that a perfect speedup is impossible by design. As spatial parallelization by domain decomposition can scale almost
perfectly until some critical number of processors where communication becomes dominant, parallel-in-time integration should be considered as a possibility for additional
fine grain parallelism on top of an existing coarse grain domain decomposition, to be used if more processors are available after the spatial parallelization has
saturated.

A second bound for $s$ emerging from \eqref{eq:speedup} is
\begin{equation}
	\label{eq:speedupBound2}
	s(N_{\rm p}) \leq \frac{ N_{\rm f} \tau_{\rm f} }{ N_{\rm c} \tau_{\rm c} } \frac{1}{1 + N_{\rm it}} \leq \frac{ N_{\rm f} }{ N_{\rm c} } \frac{ \tau_{\rm f} }{  \tau_{\rm c} }.
\end{equation}
Hence running the coarse propagator once over all coarse intervals has to be sufficiently cheaper in terms of computation time than running the fine propagator. A
reduction of the run time of the coarse propagator can be achieved either by a larger coarse time step $\Delta t$, leading to fewer coarse steps and improving the
ratio $N_{\rm f} / N_{\rm c}$ and/or by reducing the run time for a single coarse step by using a lower order scheme, lower order spatial operators, etc. and thus
improving $\tau_{\rm f} / \tau_{\rm c}$. However, note that the bounds \eqref{eq:speedupBound1} and \eqref{eq:speedupBound2} are competing in the sense that 
a cheaper, less accurate coarse propagator will improve the bound \eqref{eq:speedupBound2} but might require more iterations, thus degrading 
\eqref{eq:speedupBound1}. For the algorithm to be efficient, a reasonable balance has to be found.

The QR decomposition required in the subspace update in KSE-Parareal introduces another bound, namely
\begin{equation}
	\label{eq:qrCost}
	s(N_{\rm p}) \leq \frac{ N_{\rm f} \tau_{\rm f}}{N_{\rm it} \tau_{\rm QR}(N_{\rm it}) }.
\end{equation}
The computational complexity of the QR decomposition of a $n \times m$ rectangular matrix with $m \gg n$ is $\mathcal{O}(n^{2} m)$, see \cite{Deuflhard2002}.
In Algorithm \ref{alg:KSEParareal}, the number of columns is equal to the dimension of $\Tt{S}^{k}$ while the number of rows is equal to the number of
degrees-of-freedom $d$. The former is bounded by $N_{\rm it} N_{\rm c}$, as in each iteration the dimension of $\Tt{S}^{k}$ can increase by a maximum of $N_{\rm c}$ if all added values $\Tt{q}^{k}_{i}$, $i=0,\ldots,N_{\rm c}-1$, turn out to be linearly independent, so typically $d \gg \textrm{dim}(\Tt{S}^{k})$ and thus
\begin{equation}
 	\tau_{\rm QR}(N_{\rm it}) \sim d N_{\rm c}^2 N_{\rm it}^{2}.
\end{equation}
Evaluating $\Tt{f}$ requires at least some computations for every degree of freedom, so that an efficient implementation should yield $\tau_{\rm f} = \mathcal{O}(d)$. Hence 
\begin{equation}
	\label{eq:qrCostAsym}
	\frac{ N_{\rm f}  \tau_{\rm f} }{ N_{\rm it} \tau_{\rm QR}(N_{\rm it})} \sim \frac{ N_{\rm f} d }{N_{\rm it}  d N^{2}_{\rm it} N^{2}_{\rm c} } \sim \frac{N_{\rm f}}{N_{\rm c}  }  
	\frac{1}{N_{\rm c} N^{3}_{\rm it}},
\end{equation}
so that the computational cost for the subspace update should be independent of the problem size but will increase rapidly with the number of performed iterations. Also
\eqref{eq:qrCostAsym} results in a stricter bound for the speedup as $N_{\rm c}$ increases, even if $N_{\rm f}/N_{\rm c}$ remains constant, so performing multiple parallel steps with one coarse interval per processor will yield better speedup than fewer parallel steps where each processor handles multiple coarse intervals. 

Assuming that both the coarse and the fine propagator use the same spatial mesh, \eqref{eq:speedupBound2} can be expressed in terms of the CFL numbers of 
the two propagators, that is
\begin{equation}
	\label{eq:cflNumbers}
	C_{\rm f} = \frac{c \delta t}{\delta x}, \quad C_{\rm c} = \frac{c \Delta t}{\delta x},
\end{equation}
where $c$ is the fastest wave speed arising in a considered problem. For the acoustic-advection system in section \ref{sec:results}, $c$ is
the speed of sound. Using \eqref{eq:cflNumbers} in \eqref{eq:speedup} and neglecting the cost of the subspace update allows to derive the estimate
\begin{equation}
	\label{eq:speedupBoundEstimate}
	s(N_{\rm p}) \leq \frac{1}{ \left( 1 + N_{\rm it} \right) \frac{C_{\rm f}}{C_{\rm c}}\frac{\tau_{\rm c}}{\tau_{\rm f}} + \frac{N_{\rm p}}{N_{\rm it}}  },
\end{equation}
and it is shown in Section \ref{sec:results} that \eqref{eq:speedupBoundEstimate} gives a reasonable estimate of the actually achieved speedups in the investigated
example when using an empirically computed ratio $\tau_{\rm c} / \tau_{\rm f}$. Note that if a coarsened spatial mesh would be used for $\mathcal{G}$, an additional factor
corresponding to the ratio of the cell sizes in the coarse and fine mesh would appear in front of $C_{\rm f}/C_{\rm c}$ in \eqref{eq:speedupBoundEstimate}.

\section{Numerical Results}\label{sec:results}
In this section, first the instability of the original Parareal and the stability of KSE-Parareal for the two-dimensional advection equation
\begin{equation}
	\label{eq:advection}
	q_{t} + \Tt{U} \cdot \nabla q = 0
\end{equation}
is demonstrated. Then the performance of KSE-Parareal is addressed in detail for the linearized acoustic-advection system
{\newline
\parbox{0.35\textwidth}{\vspace{1ex}
\begin{align*}
	u_{t} + \Tt{U} \cdot \nabla u +  c_{\rm s} \pi_{x} &= 0 \\
	v_{t} + \Tt{U} \cdot \nabla v + c_{\rm s} \pi_{y} &= 0 \\
	\pi_{t} +  \Tt{U} \cdot \nabla \pi +  c_{\rm s} \left( u_{x} + v_{y} \right) &= 0,
\end{align*} }\hfill
\parbox{0.001\textwidth}{\begin{align}\label{eq:acousticAd} \end{align} } \newline }
where $c_{\rm s}$ denotes the speed of sound and $\Tt{U} = (U, V)$ a constant-in-time advection velocity. The unknowns are the perturbation velocity fields $\Tt{u} = (u, v)$ and 
the perturbation pressure $\pi$. The system \eqref{eq:acousticAd} contains two out of three major processes that govern the stability of integration schemes used for atmospheric flows, namely advection and acoustic waves. It does not include internal gravity waves whose propagation velocity ranges somewhere between typical advection speeds and the speed of sound, depending on their wave length. Note that \eqref{eq:acousticAd} is mathematically equivalent to the linearized shallow water equations.

The employed spatial discretization is a finite difference scheme in conservation form on equally sized rectangular cells. Let $q_{i}$ denote a cell centered value of some 
quantity for a cell of size $\Delta x \times \Delta y$ with index $(i,j)$. Then the rate of change at a given time is 
\begin{equation}
	\partial_{t} q_{i,j} = -\frac{F_{i+1/2,j} - F_{i-1/2,j}}{\Delta x} - \frac{G_{i,j+1/2} - G_{i,j-1/2}}{\Delta y},
\end{equation}
where $F_{i+1/2,j}$, $F_{i-1/2,j}$ are the fluxes across the interfaces in $x$-direction and $G_{i,j+1/2}$, $G_{i,j-1/2}$ the fluxes across the interfaces in
$y$-direction. The fluxes are evaluated according to the stencils of order one to six given in \cite{WickerSkamarock2002, Durran2010}. Applying
the spatial discretization to a given PDE results in an ODE system of the form \eqref{eq:ODE} which can then be solved by standard integration schemes. 
The acoustic fluxes in \eqref{eq:acousticAd} are always computed with a second order centered stencil in both the coarse and fine propagator while the order of the advective fluxes ranges between a sixth order accurate centered stencil to a first order upwind flux.

\subsection{Instability of original Parareal}
\begin{figure}[!th]
	\centering
	\includegraphics[scale=1]{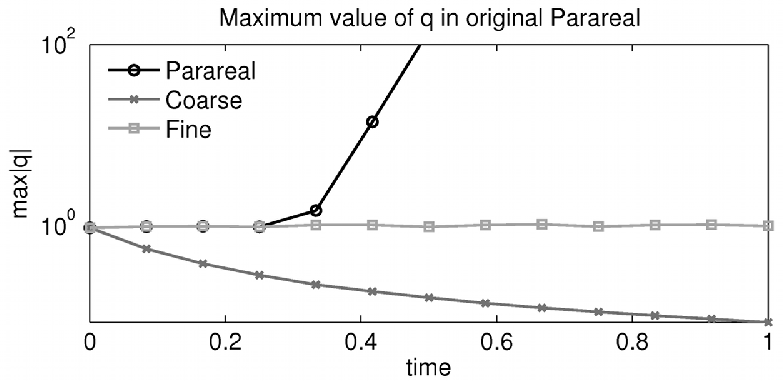}
	\includegraphics[scale=1]{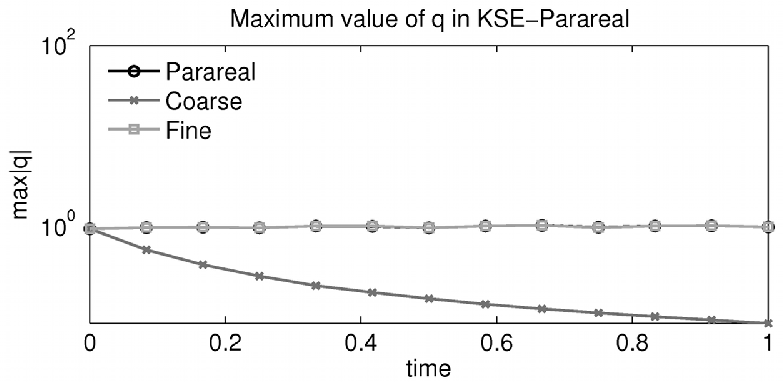}
	\caption{$\vnorm{\Tt{q}}_{\infty}$ over time for the coarse (RK-3, first order upwind flux, $C_{\rm c}=0.6$) and fine propagator (RK-3, sixth order centered flux, $C_{\rm f}=0.1$) 
	run sequentially as well as the original Parareal (upper) and KSE-Parareal (lower) for pure advection. The lines corresponding to the fine propagator and the parallel 
	solution in the lower figure do essentially coincide.}
	\label{fig:ParallelAdvection}
\end{figure}
For a brief illustration of the instability that can arise for the original version of Parareal, consider the simple two-dimensional advection problem \eqref{eq:advection}
on the unit square $[0,1] \times [0,1]$, discretized with $40 \times 40$ cells of equal size and with periodic boundary conditions. The advection velocity is set to
\begin{equation}
	\Tt{U} = (1, 1),
\end{equation}
and the simulation is run from $T=0$ to $T=1$. As initial value for $u$, a distribution also used for example in \cite{Durran2010} as a test of two-dimensional advection 
is employed,
\begin{equation}
	\label{eq:advectionIni}
	u_{0}(x,y) = \frac{1}{2} \left[ \cos\left( \pi r(x,y)  \right) +  1 \right]
\end{equation}
with
\begin{equation}
	\label{eq:defR}
	r(x,y) = \min\left(1, 4\sqrt{ \frac{(x - x_{0})^2}{0.5^2} + \frac{(y-y_{0})^{2}}{0.5^2} } \right) 
\end{equation}
and
\begin{equation}
	 x_{0}=y_{0}=0.5.
\end{equation}
The fine propagator is a Runge-Kutta-3 scheme with $C_{\rm f} = 0.1$ and sixth order fluxes, the coarse propagator is a Runge-Kutta-3 scheme with 
$C_{\rm c} = 0.6$ and first order upwind flux. The parallel algorithm uses $N_{\rm p} = 6$ coarse intervals but is run sequentially for testing purposes. $N_{\rm it} = 5$
iterations are performed in every parallel step. Figure \ref{fig:ParallelAdvection} shows $\vnorm{\Tt{q}}_{\infty}$ over time for the parallel integration scheme as well as for
the coarse and fine propagator run sequentially from $T=0$ to $T=1$. Both the fine and the coarse propagator are stable, the coarse propagator being strongly
diffusive. The solution computed by the original Parareal clearly becomes unstable at about $T=0.2$ (upper figure). Switching to the Krylov-subspace-enhanced
version (lower figure) results again in a stable scheme. 
\begin{figure}[!th]
	\centering
	\includegraphics[scale=1]{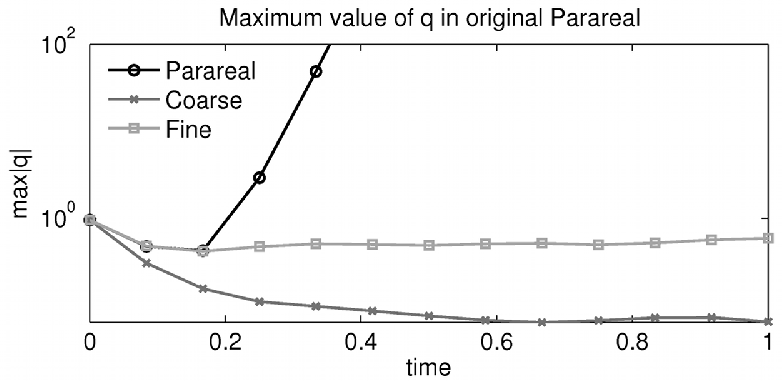}
	\includegraphics[scale=1]{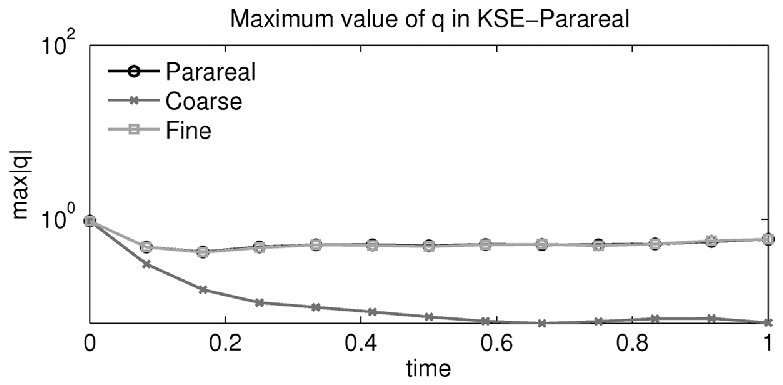}
	\caption{$\vnorm{\Tt{q}}_{\infty}$ over time for the coarse (RK-3, first order upwind flux, $C_{\rm c}=0.6$) and fine propagator (RK-3, sixth order centered flux, $C_{\rm f}=0.1$) 
	run sequentially as well as the original Parareal (upper) and KSE-Parareal (lower) for the acoustic-advection system. The lines corresponding to the fine propagator 
	and the parallel solution in the 	lower figure do essentially coincide.}
	\label{fig:ParallelAcousticAdvection}	
\end{figure}

The instability can also be demonstrated to arise for the acoustic-advection system \eqref{eq:acousticAd}. Again, use a Runge-Kutta-3 scheme with $C_{\rm f} = 0.1$ and sixth order fluxes for the advective terms for $\mathcal{F}$ and a Runge-Kutta-3 scheme with $C_{\rm c} = 0.6$ and first order upwind advective fluxes for $\mathcal{G}$. Both schemes use second order centered fluxes for the acoustic terms. The sound speed is set to $c_{\rm s} = 1.0$ for this example and \eqref{eq:advectionIni} is used as initial value for $u$ while $v$ and $\pi$ are set to zero at $T=0$. All other parameters remain as in the pure advection example above. Figure \ref{fig:ParallelAcousticAdvection} again shows the maximum absolute value for the solution vector $\Tt{q}$, containing all the components, over time. Clearly, the original Parareal is again unstable while KSE-Parareal is not.

Further tests not documented here indicate, however, that the occurrence of the instability does depend on different
factors, for example the actual schemes used for the coarse and fine propagators as well as the order of the underlying spatial discretization. The stability of Parareal applied to
semi-discrete PDEs is apparently more involved. Some analytic results concerning the stability of Parareal for PDEs can be found in \cite{Bal2003}, but for the present paper the example suffices to demonstrate the inapplicability of the original Parareal for the considered problem, confirming the requirement to use the more complex KSE-Parareal variant.

\subsection{KSE-Parareal for Acoustic-Advection}
In the following, the accuracy and efficiency of KSE-Parareal for the integration of the acoustic-advection system \eqref{eq:acousticAd} is investigated. For the 
coarse propagator $\mathcal{G}$ a partially split forward Euler scheme is used, with a forward-backward scheme for the acoustic steps. Partially split schemes
rely on the separation of the modes into fast and slow, in \eqref{eq:acousticAd} the fast modes are sound waves and the slow mode is advection. For every update
of the slow tendencies, $N_{\rm sound}$ small time steps for the fast tendencies are performed. These schemes allow for larger time steps
in the slow modes, than the stability limit enforced by the meteorologically unimportant acoustic waves permits in non-split schemes. See \cite{SkamarockKlemp1992, WickerSkamarock2002, WickerSkamarock1998, Durran2010} for details on partially split schemes. While the partially split forward Euler is found in \cite{WickerSkamarock1998} to be unstable in the semi-discrete case, it can be stabilized by using diffusive low order spatial operators and strong divergence damping, see \cite{Baldauf2010, Baldauf2002}. 
Divergence damping, introduced in \cite{SkamarockKlemp1992}, is generally required to stabilize partially split schemes, see \cite{Baldauf2010}.  Although the strong dissipation generated by $\mathcal{G}$ would be problematic when using it as a stand-alone method, it turns out that Parareal is quite effective in correcting
for the too strong damping of the coarse propagator, which is consistent with its reported good performance for diffusive problems. The momentum equations are modified to
{\newline
\parbox{0.35\textwidth}{ \vspace{1ex}
\begin{align*}
	u_{t} + \Tt{U} \cdot \nabla u + c_{\rm s} \pi_{x} &= \alpha_{1} ( \nabla \cdot \Tt{u} )_{x} \\
	v_{t} + \Tt{U} \cdot \nabla v + c_{\rm s} \pi_{y} &= \alpha_{2} (\nabla \cdot \Tt{u} )_{y},
\end{align*} }  \hfill
\parbox{0.001\textwidth}{  \begin{align} \end{align} } \newline }
with $\Tt{u} = (u, v)$ and mesh and time step dependent damping coefficients
\begin{equation}
	\label{eq:dampingCoeff}
	\alpha_{1} = \nu_{\rm c/f} \frac{\delta x^{2}}{\tau}, \ \alpha_{2} = \nu_{\rm c/f} \frac{\delta y^{2}}{\tau},
\end{equation}
with $\tau = \Delta t$ in the coarse and $\tau = \delta t$ in the fine propagator.
In the coarse propagator, when a partially split scheme is used, the coefficients become
\begin{equation}
	\alpha_{1} = \nu_{\rm c} \frac{\delta x^{2}}{\Delta t / N_{\rm sound} }, \ \alpha_{2} = \nu_{\rm c} \frac{\delta y^{2}}{\Delta t / N_{\rm sound} }
\end{equation}
as the divergence damping is employed in every acoustic step\footnote{Note that the use of a partially split scheme for $\mathcal{G}$ introduces a third time step size $\Delta t / N_{\rm sound}$ besides the two time steps of the fine and coarse propagator. No symbol is introduced for it in order to keep notation simple. The reader is cautioned, however, that $\Delta t$ and $\delta t$ refer to the time steps of $\mathcal{G}$ and $\mathcal{F}$ and not to the acoustic step size in the split scheme.}. For the fine propagator $\mathcal{F}$, a non-split Runge-Kutta-3 scheme is used. While no divergence damping is required to stabilize this scheme, it is found that when using a partially split scheme for $\mathcal{G}$, a small amount of divergence damping in $\mathcal{F}$ is required in order for the Parareal iteration to converge reasonably fast. A value of $\nu_{\rm f} = 0.005$ is used throughout this paper. The advection velocity is set to
\begin{equation}
	\label{eq:adVel}
	\Tt{U} =  \gamma \cdot \left( y - 0.5, -(x-0.5) \right), \quad \gamma = \pi,
\end{equation}
corresponding to a solid body rotation completing one full revolution in a non-dimensional time of $T=2$. The computational domain is the unit square
$[0,1] \times [0,1]$, resulting in a maximum value of $\Tt{U}$ of about $1.5$.  The sound speed is set to $c_{\rm s} = 30$, thereby roughly reproducing the typical ratio between 
advection and sound speed in atmospheric flows of about $10 \ \textrm{m} \ \textrm{s}^{-1}$ to $300 \ \textrm{m} \ \textrm{s}^{-1}$. 

Parallelization in the employed implementation is done by using OpenMP directives. All computations are performed using a computer with two sockets and one quad-core Intel Xeon processor with a frequency of $2.26 \ \textrm{GHz}$ per socket. Each socket has $8 \ \textrm{GB}$ attached and access to the memory of the other socket is slower, hence the system features "non-uniform memory access" (NUMA). Careful use of "first touch policy" is required to ensure efficiency of the multithreaded implementation, see \cite{OpenMp2008}.

The initial value for $u$ is again \eqref{eq:advectionIni} but with $y_{0} = 0.65$, that is, the maximum is not located in the center of the domain.
The $y$-component of the velocity and the pressure $\pi$ are set to zero initially. This setup is run with from $T=0$ to $T=2$ with fixed fine CFL number of
$C_{\rm f} = 0.2$ and different values for $C_{\rm c}$, $N_{\rm sound}$, $N_{\rm p}$, $N_{\rm it}$ and $\nu_{\rm f}$, $\nu_{\rm c}$ and results are presented and analyzed below.
\subsubsection{Vorticity}\label{subsubsec:vorticity}
\begin{figure*}[!th]
	\centering
	\includegraphics[scale=1]{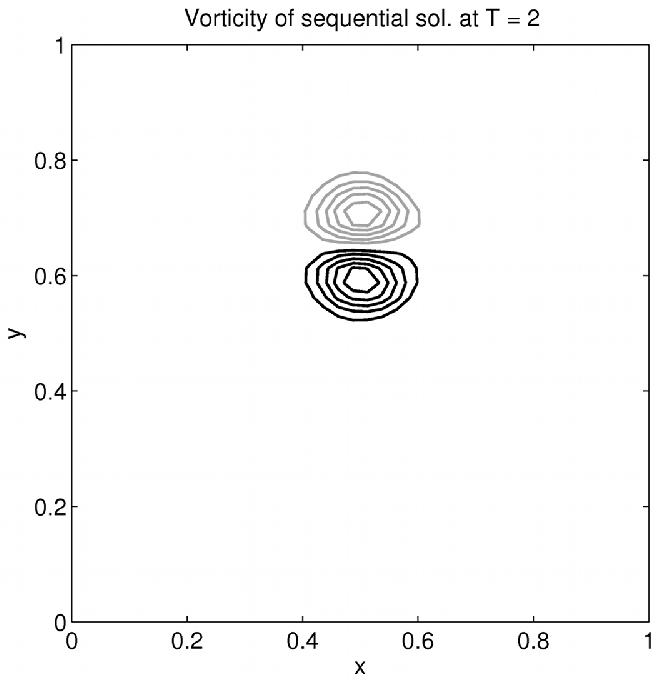}
	\includegraphics[scale=1]{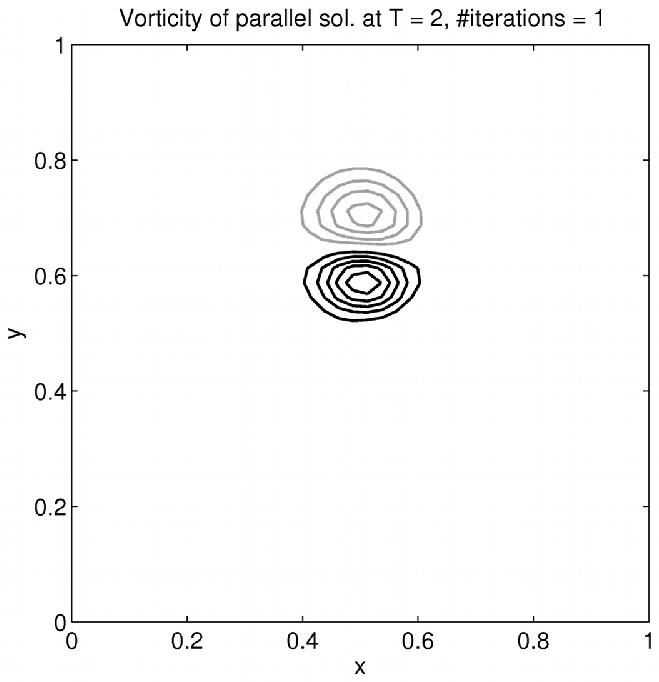}
	\includegraphics[scale=1]{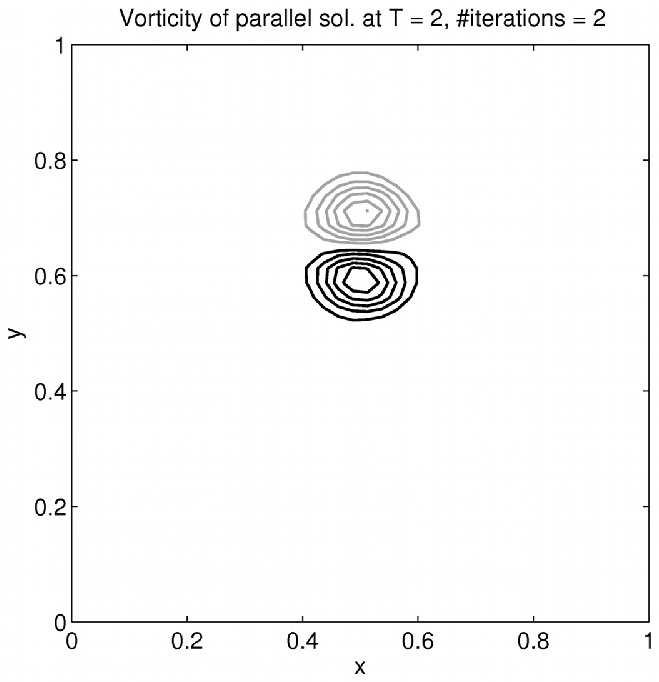}
	\includegraphics[scale=1]{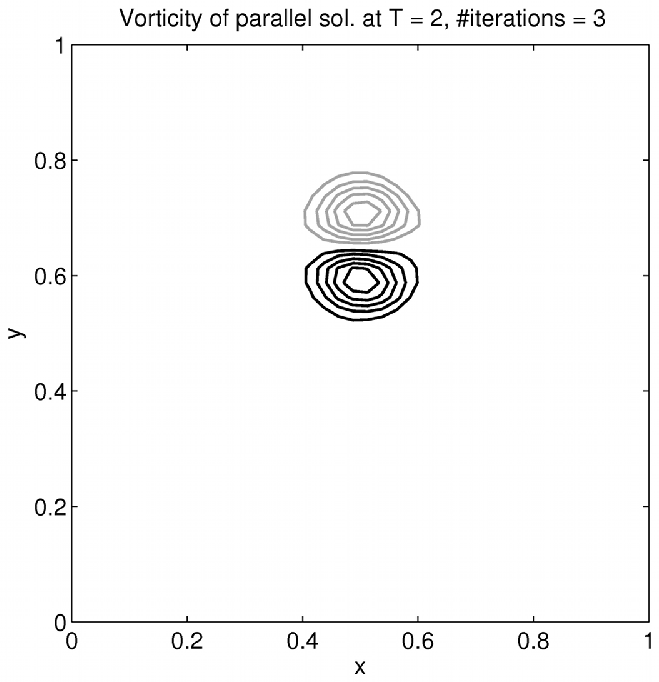}
	\caption{Vorticity of the resulting velocity field at $T=2$ after one complete revolution. The spatial resolution is $40 \times 40$ cells, the difference between isolines 
is $2$ and gray isolines correspond to negative values of vorticity. Shown are the sequentially computed reference solution (upper left) and the parallel solution for 	
$N_{\rm it}=1$ (upper right), $N_{\rm it}=2$ (lower left) and $N_{\rm it}=3$ (lower right). For all parallel solutions, a value of $N_{\rm p} = 6$ was used. The CFL number
for the coarse propagator is $C_{\rm c} = 4.0$ with $N_{\rm sound} = 4$ acoustic steps per time step. The divergence damping for the fine propagator is $\nu_{\rm f} = 0.005$ and
for the coarse propagator $\nu_{\rm c} = 0.1$. Corresponding run times can be found in Table \ref{tab:runtimesB}.}\label{fig:vorticity} 
\end{figure*} 
The two-dimensional vorticity is defined as
\begin{equation}
	\omega := u_{y} - v_{x}.
\end{equation}
Cross differentation of \eqref{eq:acousticAd} and using the special form of the advection velocity \eqref{eq:adVel} yields
\begin{equation}
	\omega_{t} + \Tt{U} \cdot \nabla \omega = \nabla \cdot \Tt{u}.
\end{equation}
The divergence at the right hand side is associated with acoustic modes and turns out to be small in the present example, so that at leading order the vorticity is simply advected
by the flow $\Tt{U}$ and has completed one full revolution at $T=2$.

Figure \ref{fig:vorticity} shows the vorticity of the obtained velocity fields at the end of the simulation for a sequentially computed reference solution and a parallel solution computed with different number of iterations per parallel time step. Even with just one iteration per parallel step, the parallel integration scheme captures the advected vorticity quite well, although some damping of the negative extremum results in a slightly asymmetric distribution. For the parallel solutions with $N_{\rm it}=2$ and $N_{\rm it}=3$, the resulting vorticity fields are indistinguishable from the reference solution. Hence KSE-Parareal not only remains stable but also does a good job in capturing the essential slow dynamics of the simulated problem, that is the advection of vorticity. Corresponding run times can be found in Tables \ref{tab:runtimesA} and \ref{tab:runtimesB}. Naturally, the solution with $N_{\rm it}=1$ yields the largest speedup, but as seen introduces noticeable errors in the vorticity field. Parallel solutions with two or three iterations, both showing very good agreement of the resulting vorticity fields with the reference solution, still allow for a reduction of the time-to-solution by a factor of $1.9$ and $1.3$ respectively.

\subsubsection{Time Series}
\begin{figure}[!th]
	\centering
	\includegraphics[scale=1]{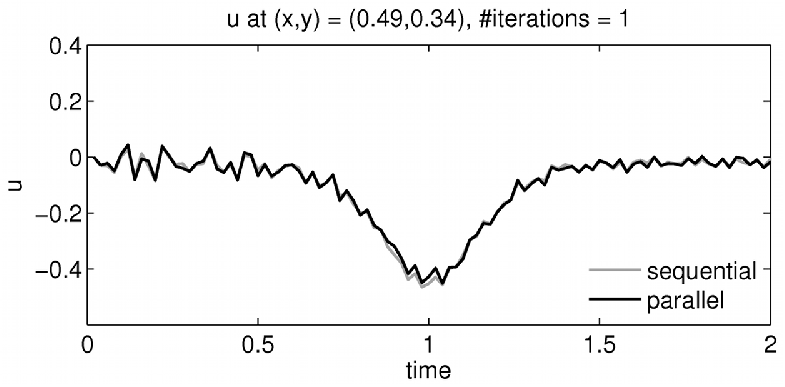}
	\includegraphics[scale=1]{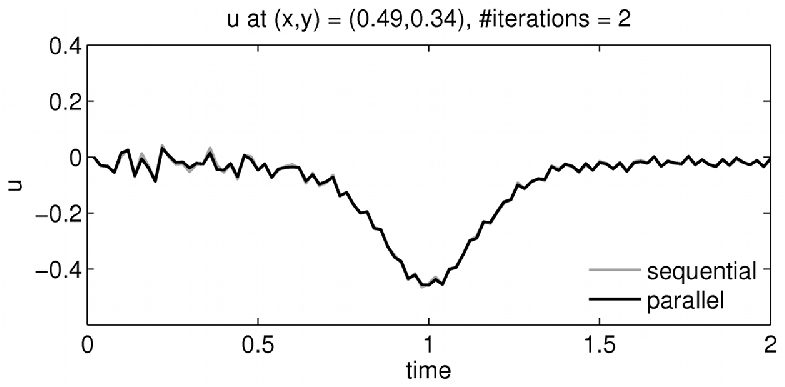}	
	\caption{Horizontal velocity $u$ at $(x,y) = (0.49, 0.34)$ over time, plotted after each completed parallel step. The solid line represents the parallel solution
	after one (upper) or two (lower) iteration while the dashed, gray line represents the sequentially computed reference solution. The simulation parameters are
	identical to the ones given for Figure \ref{fig:vorticity}.}\label{fig:UoverTime}
\end{figure}

\begin{figure}[!th]
	\centering
	\includegraphics[scale=1]{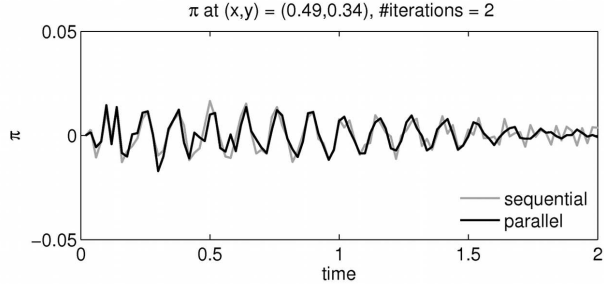}
	\caption{Pressure $\pi$ at $(x,y) = (0.49, 0.34)$ over time, plotted after each completed parallel step. The solid line represents the parallel solution
	after one (left) or two (right) iteration while the dashed, gray line represents the sequentially computed reference solution. The simulation parameters are
	identical to the ones given for Figure \ref{fig:vorticity}.}\label{fig:PoverTime}
\end{figure}

Figure \ref{fig:UoverTime} shows plots of the horizontal velocity of the parallel computed solution for $N_{\rm it}=1$ and $N_{\rm it} = 2$ at $(x,y) = (0.49, 0.34)$, that is, the
point opposite to the center of the initial distribution. The agreement in the horizontal velocity is very good, even if only one iteration per parallel step is used. Some small
differences are visible in the minimum passing through the point at $T=1$. For $N_{\rm it} = 2$, the parallel and sequential solution are basically indistinguishable. The
representation of the pressure $\pi$, shown in figure \ref{fig:PoverTime} is still surprisingly good, given that the partially split scheme employed $\mathcal{G}$ does by
design not represent sound waves very accurately. Increasing the number of iterations provides a more accurate resolution of the acoustic modes, but also severely 
inhibits the efficiency of the parallel scheme. For a highly accurate representation of the acoustic modes with a small number of iterations, a non-split scheme for $\mathcal{G}$ 
is most likely required, which would however reduce the efficiency of the time-parallel scheme.
	
\subsubsection{Energy}\label{subsubsec:energy}
\begin{figure}[!th]
	\centering
	\includegraphics[scale=1]{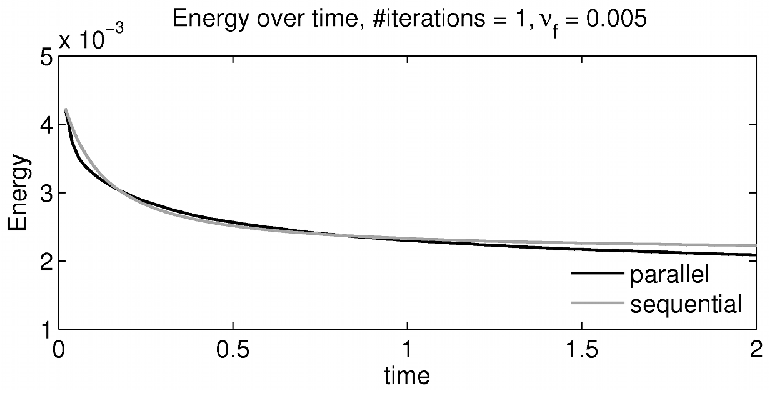}
	\includegraphics[scale=1]{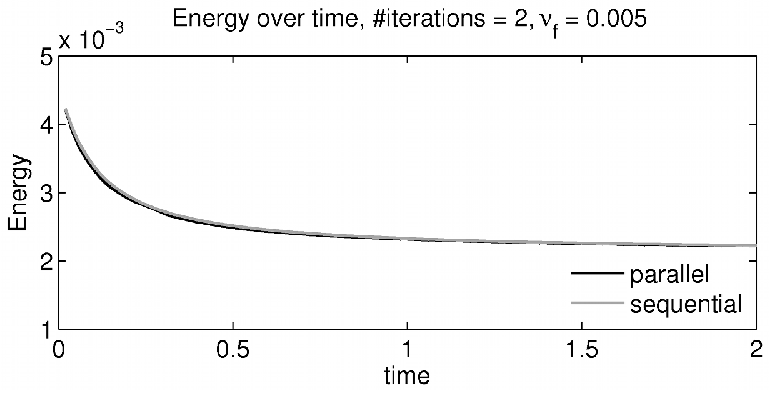}
	\caption{Total energy over time for the sequential solution and the parallel solution with $N_{\rm it}=1$ (upper) and $N_{\rm it} = 2$ (lower). The parameters
	of the simulation are identical to the ones in Figure \ref{fig:vorticity}. Note the divergence damping removes energy associated with acoustic waves, so
	the total energy over time decreases.}\label{fig:EoverTime1}
\end{figure}

The system \eqref{eq:acousticAd} allows for the definition of the energy density
\begin{equation}
	E = \left( u^{2} + v^{2} + \pi^{2} \right).
\end{equation}
By basic manipulations of \eqref{eq:acousticAd} one can derive
\begin{equation}
	E_{t} + \Tt{U} \cdot \nabla E + c_{\rm s} \nabla \cdot E = 0.
\end{equation}
Integrating over the domain, using the divergence theorem as well as the assumed periodicity of the boundary values yields that the total energy
\begin{equation}
	E_{\rm tot} = \int_{ [0,1]^{2} } E \ d\Tt{x}
\end{equation}
is conserved in the original system. However, because of the introduced divergence damping, this is no longer true for the numerical solution. Figure \ref{fig:EoverTime1}
shows a comparison of the evolution of the total energy for the parallel and the sequential solution. Even for the sequential solution, the divergence damping
results in a significant reduction of the total energy, although the damping parameter $\nu_{\rm f} = 0.005$ is quite small. The parallel solution computed with a single
iteration per parallel step is slightly more diffusive than the sequentially computed, as apparently one iteration is not enough to correct for the strong diffusivity
of the coarse propagator. Performing two iterations per parallel step yields an energy evolution that is indistinguishable from the one of the sequential solution.

\begin{figure}[!th]
	\centering
	\includegraphics[scale=1]{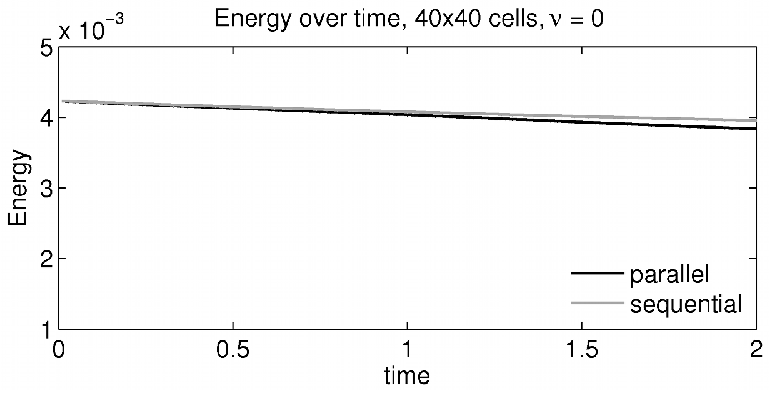}
	\includegraphics[scale=1]{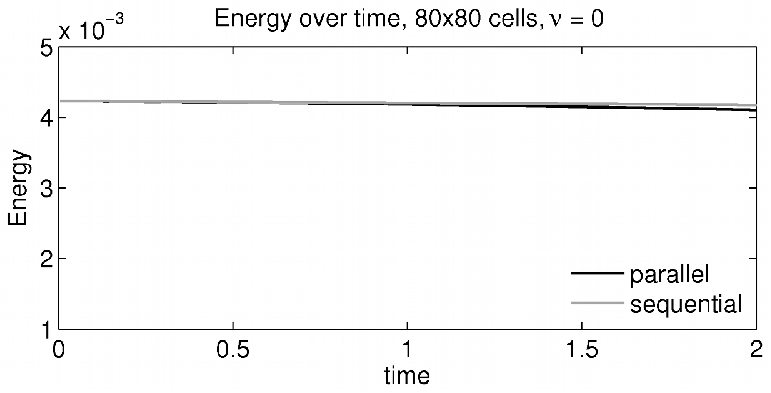}
	\caption{Total energy over time on a mesh with $40 \times 40$ cells (upper) and $80 \times 80$ cells (lower). There is no divergence damping for the
	fine propagator, that is $\nu_{\rm f} = 0$, and the divergence damping for the coarse propagator is reduced to $\nu_{\rm c} = 0.005$. Also, coarse CFL number
	is reduced to $C_{\rm c} = 2.0$ and the massively diffusive
	first order upwind flux for the coarse propagator has been replaced by a somewhat less diffusive third order flux. To maintain stability, the number of
	acoustic steps in the coarse scheme has been increased to $N_{\rm sound} = 8$. In both cases, the parallel computation used $N_{\rm p} = 6$.}\label{fig:EoverTime2}
\end{figure}

In order to assess how the parallel integrator performs in terms of energy conservation if no divergence damping is present in the fine integrator and a much less
diffusive $\mathcal{G}$ is used, an additional simulation is run with $\nu_{\rm f} = 0$. Also, the divergence damping in the coarse propagator is reduced to a value of 
$\nu_{\rm c} = 0.005$, the CFL number reduced to $C_{\rm c}=2$ and the first order advective flux is replaced by a less diffusive third order flux. To stabilize the coarse scheme, an increase of the number of acoustic steps to $N_{\rm sound} = 8$ is necessary. Figure \ref{fig:EoverTime2} shows the total energy over time for a mesh of $40 \times 40$ cells and also for the solution computed on refined mesh featuring $80 \times 80$ cells. The energy defect at the end of the simulation is now down to $3 \ \%$ on the coarse and $2 \ \%$ on the fine mesh. Hence using a less diffusive $\mathcal{G}$ leads, as can be expected, to a less diffusive parallel solution. However, this setup yields only small speedups and if better
conservation properties are desired, using a different scheme for $\mathcal{G}$ is most likely in order.
\subsubsection{Runtimes and Speedup}
All run times indicated below are measured using the OpenMP function OMP\_GET\_WTIME and do not include I/O operations. No averaging over an ensemble of different runs was performed, but the run times proved to be very stable with variations only on the order of a few percent over different runs, so that the indicated values are representative.

\begin{figure*}[t]
	\centering
	\includegraphics[scale=1]{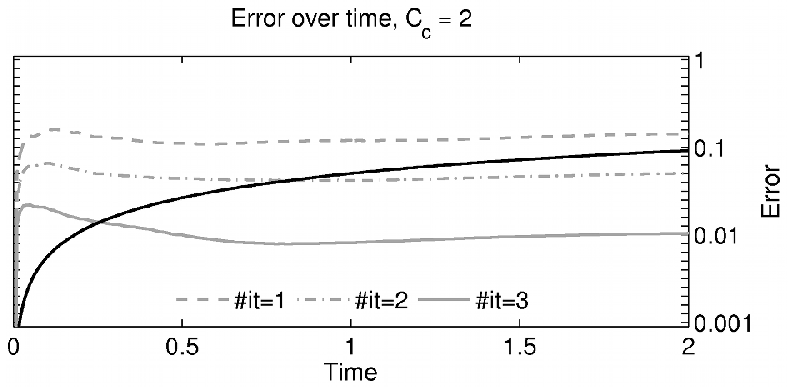}
	\includegraphics[scale=1]{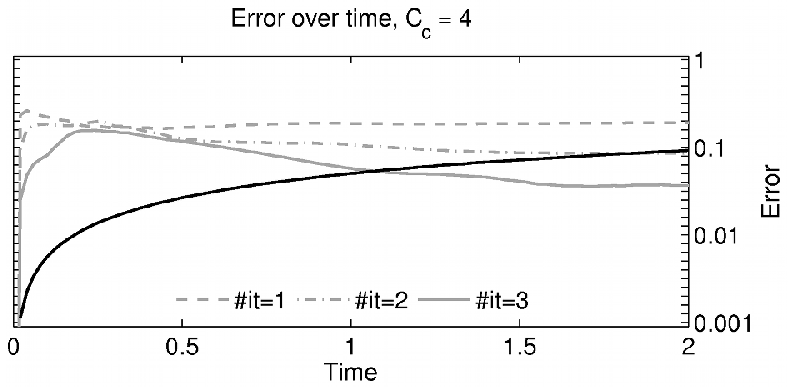}
	\caption{Difference between the parallel and the sequential solution for $N_{\rm it} = 1,2,3$ (grey lines) for $C_{\rm c} = 2.0$ (upper) and $C_{\rm c} = 4.0$ (lower) and 
	estimated temporal discretization error of $\mathcal{F}$ (black line) over time. The temporal discretization error is estimated by comparing the solution obtained by running 
	$\mathcal{F}$ sequentially to a reference solution obtained using a ten times smaller time step while disabling divergence damping in both runs. The other parameters of the   parallel runs are as indicated in Figure \ref{fig:vorticity}.}\label{fig:ErrorVsTime}
\end{figure*}

Tables \ref{tab:runtimesA} and \ref{tab:runtimesB} show errors, run times, speedup and parallel efficiency for parallel solutions computed for coarse CFL numbers of 
$C_{\rm c}=2$ and $C_{\rm c}=4$ with different numbers of iterations using $N_{\rm p} = 4, 6, 8$ processors. The error indicates the relative $l_{2}$-norm of the
difference between the solution $\Tt{q}_{\rm seq}$ obtained by running $\mathcal{F}_{\delta t}$ sequentially and the solution of the parallel scheme $\Tt{q}_{\rm par}$,
that is
\begin{equation}
	\label{eq:errorDef}
	\epsilon := \frac{\vnorm{\Tt{q}_{\rm par} - \Tt{q}_{\rm seq} }_{2}}{\vnorm{\Tt{q}_{\rm seq}}_{2}}.
\end{equation}
Tables \ref{tab:runtimesA}, \ref{tab:runtimesB} show $\epsilon$ evaluated at the end of the simulation while Figure \ref{fig:ErrorVsTime} shows $\epsilon$ evaluated
after every parallel step. Speedup is the ratio of the run time of the reference solution to the run time of the parallel solution and efficiency is the speedup divided by the number of used processors. In all runs, the number of coarse intervals per parallel step is equal to the number of processors so that runs with smaller values of $N_{\rm p}$ perform more but shorter parallel steps while those with larger values of $N_{\rm p}$ perform fewer larger steps.

In general, every iteration reduces the error by about half an order of magnitude, exceptions being the cases in Table \ref{tab:runtimesB} for $N_{\rm p} = 4$, $N_{\rm it}=2$
and $N_{\rm p}=8$, $N_{\rm it}=4$ where only a marginal reduction of the error is observed. As expected from the theoretical analysis in subsection 
\ref{subsec:expectedSpeedup}, achieved speedups depend very strongly on the number of iterations performed and noticeable acceleration requires the algorithm to 
terminate after two or at most three iterations. Comparing the values in the "error" as well as the "speedup" columns in \ref{tab:runtimesA} and \ref{tab:runtimesB} reveals that a larger value of $C_{\rm c}$ noticeably improves the obtained speedup at the expense of a less accurate solution. 

Note, however, that there is no real benefit from approximating the sequential solution with an accuracy that is higher than the discretization error of the sequential solution. In
order to estimate the temporal discretization error, the fine propagator is run with a ten times smaller time step and the difference between $\mathcal{F}_{\delta t}$ and
$\mathcal{F}_{\delta t / 10}$ is used as an estimate for the error of the temporal discretization. As the divergence damping inhibits proper convergence of the RK-3 scheme,
this estimate is generated by running both $\mathcal{F}$ with $\nu_{\rm f} = 0$, so that a fair comparison is obtained. Figure \ref{fig:ErrorVsTime} shows the difference measured
according to \eqref{eq:errorDef} between the two sequential solutions, $\mathcal{F}_{\delta t}$ and $\mathcal{F}_{\delta / 10}$, without damping as well as the difference between the parallel and the $\mathcal{F}_{\delta t}$-sequential solution with $\nu_{\rm f} = 0.005$ for coarse CFL numbers of $C_{\rm c} = 2.0$ (upper) and $C_{\rm c} = 4.0$ (lower). For
$N_{\rm it} = 2$, at the end of the simulation the estimated discretization error is about equal to the iteration error of the parallel solution for $C_{\rm c} = 4.0$ and noticeably
smaller for $C_{\rm c} = 2.0$. Tables \ref{tab:runtimesA} and \ref{tab:runtimesB} indicate that for $N_{\rm it} = 2$, depending on the number of processors,  speedups ranging from $1.2$ to $2.1$ can be obtained. There is an interesting possibility of coupling Parareal with adaptive time step refinement here: As an
adaptive scheme has to provide estimates of the discretization error anyhow, this estimate could be reused to adaptively modify the prescribed error tolerance for the Parareal
iteration if a stopping criterion is used instead of a fixed iteration number.
\begin{table}[!th]
	\centering
	\small
	\begin{tabular}{|c|c|c|c|c|}
		\hline
		\multicolumn{5}{|c|}{$N_{\rm p} = 4$, $C_{\rm c} = 2.0$} \\ \hline
		 $\# it.$ & Error & Runtime & Speedup & Efficiency \\ \hline
		 $1$ & $1.3 \times 10^{-1}$ & $  7.4 \ \textrm{s}$ & $2.0$ & 50 \% \\
		 $2$ & $2.8 \times 10^{-2}$ & $13.0 \ \textrm{s}$ & $1.2$ & 29 \% \\ \hline
	\end{tabular}	
	\hspace*{1em}
	\begin{tabular}{|c|c|c|c|c|}
		\hline
		\multicolumn{5}{|c|}{$N_{\rm p} = 6$, $C_{\rm c} = 2.0$} \\ \hline
		 $\# it.$ & Error & Runtime & Speedup  & Efficiency \\ \hline
		 $1$ & $1.4 \times 10^{-1}$ & $  6.3 \ \textrm{s}$ & $2.4$ & 40 \% \\
		 $2$ & $5.0 \times 10^{-2}$ & $10.9 \ \textrm{s}$ & $1.4$ & 23 \% \\
		 $3$ & $1.1 \times 10^{-2}$ & $15.5 \ \textrm{s}$ & $1.0$ & -- \% \\ \hline
	\end{tabular}	
	\vspace*{1em}
	\begin{tabular}{|c|c|c|c|c|}
		\hline
		\multicolumn{5}{|c|}{$N_{\rm p} = 8$, $C_{\rm c} = 2.0$} \\ \hline
		 $\# it.$ & Error & Runtime & Speedup & Efficiency \\ \hline
		 $1$ & $1.6 \times 10^{-1}$ & $  5.8 \ \textrm{s}$ & $2.6$ & 32 \% \\
		 $2$ & $6.7 \times 10^{-2}$ & $10.1 \ \textrm{s}$ & $1.5$ & 19 \%  \\
		 $3$ & $1.7 \times 10^{-2}$ & $14.7 \ \textrm{s}$ & $1.0$ & --  \\ 
		 $4$ & $3.2 \times 10^{-3}$ & $19.6 \ \textrm{s}$ & $0.8$ & -- \\ \hline		 
	\end{tabular}	
	\caption{Results for a mesh using $40 \times 40$ cells, $C_{\rm fine} = 0.2$, $N_{\rm sound}=4$ and $C_{\rm coarse} = 2.0$  for
$N_{\rm p} = 4$ (upper), $N_{\rm p} = 6$ (middle) and $N_{\rm p} = 8$ (lower) coarse intervals per parallel time step distributed on an equal amount of processors.
The divergence damping for the fine propagator is set to $\nu_{\rm f} = 0.005$ while the coarse propagator used $\nu_{\rm c} = 0.1$. The column "Error" indicates the difference \eqref{eq:errorDef} between the parallel solution and the sequentially computed reference solution in all three fields, $u$, $v$ and $\pi$, at $T=2$ after one complete revolution. The run time for the sequential integration is $14.9 \ \textrm{s}$.}\label{tab:runtimesA}
\end{table}
\begin{table}[t]
	\centering
	\small
	\begin{tabular}{|c|c|c|c|c|}
		\hline
		\multicolumn{5}{|c|}{$N_{\rm p} = 4$, $C_{\rm c} = 4.0$} \\ \hline
		 $\# it.$ & Error & Runtime & Speedup & Efficiency \\ \hline
		 $1$ &$1.8 \times 10^{-1}$ & $  5.6 \ \textrm{s}$ & $2.7$ & 68 \% \\
		 $2$ &$1.7 \times 10^{-1}$ & $10.1 \ \textrm{s}$ & $1.5$ & 37 \% \\ \hline
	\end{tabular}	
	\hspace*{1em}
	\begin{tabular}{|c|c|c|c|c|}
		\hline
		\multicolumn{5}{|c|}{$N_{\rm p} = 6$, $C_{\rm c} = 4.0$} \\ \hline
		 $\# it.$ & Error & Runtime & Speedup & Efficiency \\ \hline
		 $1$ & $1.9 \times 10^{-1}$ & $  4.5 \ \textrm{s}$ & $3.4$ & 56 \% \\
		 $2$ & $8.6 \times 10^{-2}$ & $  8.0 \ \textrm{s}$ & $1.9$ & 31 \% \\
		 $3$ & $3.7 \times 10^{-2}$ & $11.6 \ \textrm{s}$ & $1.3$ & 22 \% \\ \hline
	\end{tabular}	
	\vspace*{1em}
	\begin{tabular}{|c|c|c|c|c|}
		\hline
		\multicolumn{5}{|c|}{$N_{\rm p} = 8$, $C_{\rm c} = 4.0$} \\ \hline
		 $\# it.$ & Error & Runtime & Speedup & Efficiency \\ \hline
		 $1$ & $2.0 \times 10^{-1}$ & $  3.9 \ \textrm{s}$& $3.9$ & 48 \% \\
		 $2$ & $9.3 \times 10^{-2}$ & $  7.9 \ \textrm{s}$& $2.1$ & 27 \% \\
		 $3$ & $3.4 \times 10^{-2}$ & $10.2 \ \textrm{s}$& $1.5$ & 18 \% \\ 
		 $4$ & $3.0 \times 10^{-2}$ & $13.6 \ \textrm{s}$& $1.1$ & 14 \% \\ \hline		 
	\end{tabular}		
	\caption{Results for simulations with the same parameters as in Table \ref{tab:runtimesA}, but now with a coarse CFL number $C_{\rm c} = 4.0$, again
	for $N_{\rm p} = 4$ (upper), $N_{\rm p} = 6$ (middle) and $N_{\rm p}=8$ (lower) processors.}	\label{tab:runtimesB}
\end{table}

\begin{figure*}[t]
	\centering
	\includegraphics[scale=1]{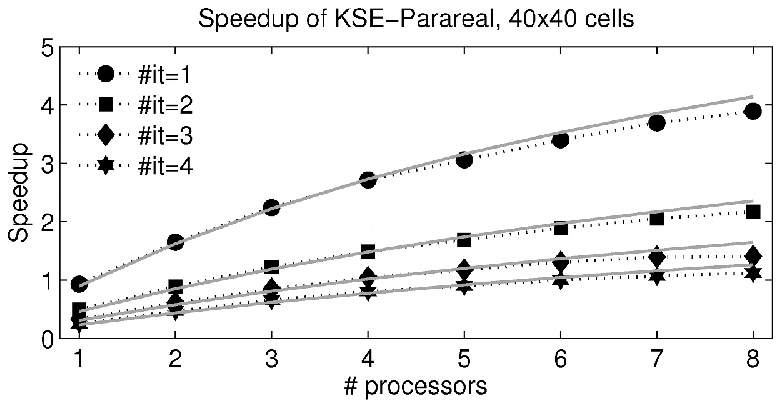}
	\includegraphics[scale=1]{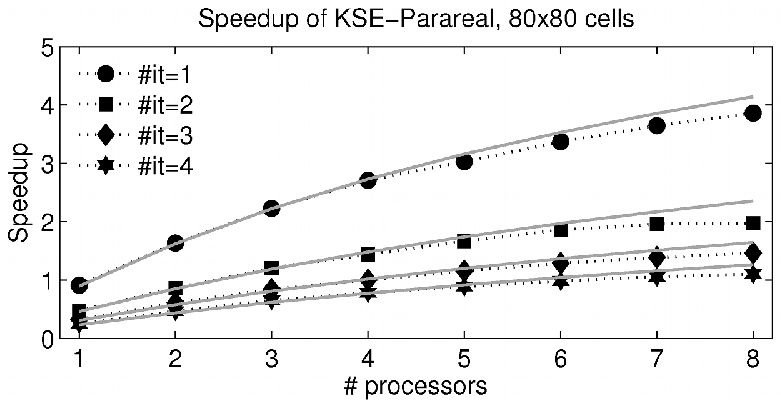}
	\caption{Measured speedup depending on the number of processors for $N_{\rm it} = 1, 2, 3, 4$. The coarse CFL number is $C_{\rm c} = 4.0$, using 
	$N_{\rm sound}=4$ acoustic steps in every large step. The mesh consists of $40 \times 40$ cells (left) or $80 \times 80$ cells (right). 
	All other parameters are as indicated in Figure \ref{fig:vorticity}. The grey lines show the speedup estimate obtained by \eqref{eq:speedupBoundEstimate} 
	with an empirically determined ratio $\tau_{\rm f} / \tau_{\rm c}$.}\label{fig:speedup}
\end{figure*}

A more complete study of the obtained speedup is found in Figure \ref{fig:speedup}. It shows the achieved speedup for a number of processors ranging between 
one and eight for parallel solutions performing $N_{\rm it} = 1, 2, 3, 4$ iterations. The upper figure shows simulations on a grid consisting of $40 \times 40$ cells, the
lower figure on a $80 \times 80$ cell grid. Again, the speedup depends critically on the number of iterations and the simulations with four iterations produce nearly no
speedup at all, even if using eight processors. As a consequence of the inherently moderate scaling, gains in speedup quickly deteriorate as the number of processors
increases, again advocating the use of Parareal for adding fine grain parallelism inside nodes. However, solutions with $N_{\rm it} = 2$, providing reasonable accuracy, 
still allow for an acceleration of about a factor two if using six processors. 

The grey lines in Figure \ref{fig:speedup} indicate the estimated speedup according to \eqref{eq:speedupBoundEstimate} with the ratio $\tau_{\rm c} / \tau_{\rm f}$ being
computed from sequential test runs. As the partially split scheme requires several acoustic steps, a single coarse step is even slightly more expensive than a single
step of $\mathcal{F}$, resulting in a ratio of $\tau_{\rm c} / \tau_{\rm f} \approx 1.165$. The reason is that evaluating the discrete advective flux is relatively straightforward in the linear case and requires neither averaging of velocities to cell interfaces nor any form of limiting. Including both procedures would increase the computational cost of evaluating
the advective tendencies, reduce the ratio $\tau_{\rm c} / \tau_{\rm f}$ and improve the parallel scaling. Also, an implementation coupling the space and time
discretization more closely than the here used method-of-lines approach might allow for a more efficient implementation of the partially split scheme. Note that estimate 
\eqref{eq:speedupBoundEstimate} neglects the run time required for the subspace update, but nevertheless gives a quite good estimate of the speedups actually obtained, supporting the point that in the presented examples the time required for the QR decomposition plays only a minor role. 

For comparison, a partially split Runge-Kutta-3 / for\-ward-back\-ward scheme is run sequentially to integrate the test problem with $C_{\rm sound} = 1.2$ and $N_{\rm sound} = 6$ as well as $C_{\rm sound} = 2.4$ and $N_{\rm sound} = 12$ and, as before, a divergence damping of $\nu_{\rm f} = 0.005$, an end time of $T=2$ and a mesh with $40 \times 40$ cells.  Switching to the partially split scheme reduces the run time to $8.3$ seconds for $C_{\rm sound} = 1.2$ and further to $7.3$ seconds for $C_{\rm sound} = 2.4$, corresponding to a speedups of $1.8$ and $2.0$ compared to the non-split version. The relative $l_{2}$-differences \eqref{eq:errorDef} between the split and non-split solutions in both cases is about $1.6 \times 10^{-1}$. Comparing this with the results in Table \ref{tab:runtimesA} shows that KSE-Parareal achieves this level of accuracy with only a single iteration and provides larger speedups in this case. As mentioned above, the used method-of-lines approach is not optimal for the partially split scheme and a properly tuned
implementation might yield runtimes comparable to or better than the parallel scheme. An even larger reduction of the overall time-to-solution of KSE-Parareal could be achieved by using a split scheme for $\mathcal{F}$, too. However, one would have to deal with two sources of error in this case: the splitting error as well as the error from the iteration of the parallel scheme.

\begin{figure}[!th]
	\centering
	\includegraphics[scale=1]{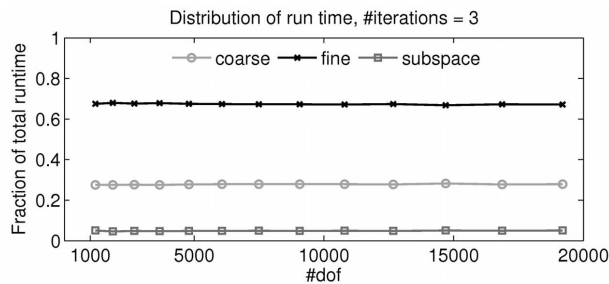}
	\caption{Fraction of run time (wall clock time) spent in the coarse propagator, the fine propagator and the subspace update for increasingly fine meshes, that
	is a growing number of degrees-of-freedom. All other parameters are as indicated in Figure \ref{fig:vorticity}.}\label{fig:percentageRuntime}
\end{figure}

Finally, Figure \ref{fig:percentageRuntime} shows the distribution of the total run time of the parallel scheme over the three parts of the iteration, that is the coarse and fine propagator and the subspace update, depending on the total number of degrees-of-freedom. The run times are wall clock times, that is the run time for the fine propagator corresponds to the time required to compute the fine integration step \emph{in parallel}. The parameters, aside from the number of cells, are as indicated in Figure \ref{fig:vorticity}. 
As it is to be expected from the asymptotic estimate \eqref{eq:qrCostAsym}, the fraction of the total run time spent in the subspace update does not depend on the
problem size. For the case with $N_{\rm it} = 3$ iterations shown in Figure \ref{fig:percentageRuntime},  about  $5 \ \%$ of the total time are spent for the subspace update, so the overhead for the Krylov-subspace-enhanced version of Parareal compared to the original version is moderate, even when computing a full QR decomposition in every iteration.

\section{Discussion}
The paper addresses the applicability of the Krylov-subspace-enhanced Para\-real parallel-in-time in\-te\-gra\-tion scheme to a two-di\-men\-sional, hyper\-bolic, linear acoustic-advection problem, which is often used as a test problem for in\-te\-gra\-tion methods for numerical weather prediction. First it is shown that the original version of Parareal, as expected from the results of studies by other authors, can become unstable, even if used to integrate a comparatively simple two-dimensional advection problem. The KSE-Parareal variant, by contrast, does not suffer from this instability. As KSE-Parareal has up to now not been applied to hyperbolic flow problems, this was not necessarily to be expected. In contrast to the original version of Parareal, the KSE variant requires a QR decomposition in every iteration. The results in the present paper show that this can be done by using an implementation of the LAPACK library in a straightforward way without severely increasing execution times.

The possibility of using a partially split scheme as coarse propagator and thus avoiding the necessity of implementing an implicit method is established. However, in order for
the Parareal iteration to converge reasonably fast, a small amount of divergence damping is required not only in the coarse but also in the non-split fine propagator.
It is demonstrated that the parallel-in-time scheme can capture important features like advection of vorticity and the amount of diffusivity of the parallel scheme in comparison to the underlying sequential scheme is discussed. Obtained speedups are indicated and it is found that reasonably accurate solutions can be computed while still allowing for accelerations in the range of factors of $1.5$ to $2.1$ using between four and eight CPUs. Comparing the speedups achieved by KSE-Parareal with the acceleration obtained by switching to a partially split scheme shows that KSE-Parareal provides larger speedup while producing solutions of comparable accuracy, but the implementation of the partially
split scheme is probably not the most efficient.

Implementing Parareal essentially requires the iterated application of two integration schemes. As the numerical examples employ integration schemes that are already widely used in contemporary codes for modeling atmospheric flows (Runge-Kutta-3, forward-backward, forward Euler), a possible future implementation into an existing code could likely rely heavily on already implemented features.

The successful application of KSE-Parareal to a nontrivial, meteorologically interesting test problem demonstrated in this paper indicates that parallel-in-time schemes are promising candidates for increasing the level of parallelism in codes for numerical weather prediction beyond the already employed spatial parallelization. Such increases will be necessary to achieve a maximized reduction of the time-to-solution on existing and, probably more important, on upcoming high performance computing systems featuring a rapidly growing number of CPUs. Also, it is stressed that the Parareal algorithm allows for parallel computation by design and is not bound to a specific model of parallelization, so its usefulness is not dependent on a specific architecture. Hence parallel-in-time schemes increase parallelism on the algorithmic side, thus potentially offering lasting benefits beyond tuning an existing implementation.

Important issues that have to be addressed in future works include: (i) Addressing the performance of the scheme to a model of increased complexity, for example the
compressible Boussinesq system. While the parallel scaling will most likely benefit from a computationally heavier problem, the ability of the scheme to produce accurate solutions within a few iterations has to be demonstrated. (ii) Evaluating the performance of a combined spatial/temporal parallelization. Because of the limited scaling of the parallel-in-time scheme, the most promising approach seems the use of a hybrid MPI/OpenMP parallelization, where subdomains are assigned to computational nodes that communicate by message passing while the time stepping within the subdomains is done in parallel using the shared memory and the CPUs available inside a node. It remains to be demonstrated that such a hybrid approach can reduce the time-to-solution below what a pure spatial parallelization can achieve. (iii) The extension of the Krylov-subspace enhancement for PITA
introduced in \cite{FarhatCortial2008} relies on the use of an implicit scheme. An adaption for explicit schemes would be desirable in order to avoid the necessity
of implementing an implicit scheme into existing NPW codes. (iv) Investigating alternatives to the used integration schemes with respect to their performance in the Parareal framework. A simple modification would be to use two partially split schemes for both, the coarse and the fine propagator, in order to further reduce run times. But also completely different schemes might be worth considering.  As examples, improved split-explicit schemes as introduced in \cite{Gassmann2005} or \cite{WenschEtAl2009} are mentioned. Interesting candidates are also the explicit peer-methods introduced in \cite{JebensEtAl2009}, that allow for a splitting between low and high frequency modes without requiring  artificial damping. Exploring the possibility of using a reduced-physics sound-proof model in the coarse propagator might also be of interest.

\section*{Acknowledgments}
This work is funded by the Swiss "High Performance and High Productivity Computing Initiative" (HP2C). The implementation uses the "Automatically Tuned Linear Algebra
Software" (ATLAS)\footnote{\url{www.math-atlas.sourceforge.net}}, the GCC compiler suit, the FORTRAN Namelist Suite for MATLAB by S. Lazerson and  the NetCDF library\footnote{\url{www.unidata.ucar.edu/software/netcdf/}}.

\bibliographystyle{model3-num-names}
\bibliography{Parareal,HPC,NWPTimeIntegration}







\end{document}